\documentclass{article}
\usepackage{mrs2005,epsfig}

\newcommand{\msun}{\,\hbox{M$_{\odot}$}}
\newcommand{\kms}{\,\hbox{\hbox{km}\,\hbox{s}$^{-1}$}}
\newcommand{\vrot}{\,\hbox{$V_{\rm rot}$}}
\newcommand{\mbh}{\,\hbox{M$_{\rm BH}$}}
\newcommand{\msigma}{\,\hbox{M$_{\rm BH}-\sigma$}}

\begin{document}

\title{Evolution of local Ultraluminous mergers from NIR spectroscopy}
 
\author{K. M. Dasyra, L. J. Tacconi, R. I. Davies, R. Genzel, D. Lutz,}
\affil{Max Planck Institut f\"ur Extraterrestrische Physik,
Postfach 1312, 85741, Garching, Germany}

\author{T. Naab, and A. Burkert}
\affil{University Observatory Munich, Scheinerstrasse 1, 81679, Munich, 
Germany}

\begin{abstract}
We present results from our VLT Large Program to study the dynamical evolution
of Ultraluminous Infrared Galaxies (ULIRGs) which are the products of mergers
of gas-rich galaxies. We have so far obtained near-infrared high-resolution 
ISAAC spectra of 53 local ULIRGs at several merger timescales and 12 
Palomar-Green 
QSOs (more than half of which are IR-bright sources). We have extracted 
the stellar velocity dispersion $\sigma$ and rotational velocity $V_{rot}$ 
along our slits to derive the kinematics of the merging galaxies.
These quantities enable us to answer the following questions about the 
evolution of ULIRGs:
1) What are the progenitor mass ratios?, 2) How do the stellar kinematics
evolve?, and, 3) Is there a connection between ULIRGs and QSOs? We find that 
the Ultraluminous phase is mainly triggered by mergers of approximately equal 
mass galaxies, however, less violent minor mergers (of progenitor mass ratio 
~3:1) also exist in our sample. Dynamical heating of the merging hosts is 
observed as the stellar systematic rotation decreases with time in favour of 
the increase
of random motions. The merger remnants, being dispersion-dominated systems
with non-negligible rotation (\vrot/sigma $\sim$ 0.6), resemble elliptical
galaxies \cite{genzel01}. Placing ULIRGs on the fundamental plane of 
early-type galaxies \cite{djoda} shows that they resemble intermediate-mass,
disky-isophote-profile ellipticals. After the nuclear coalescence, 
the black hole masses of ULIRGs, calculated from their relation to the host 
dispersions, are of the order 10$^7$ - 10$^8$\msun, thus, they do not 
resemble the supermassive BHs found in local QSOs that are selected to be
radio loud (or to have radio loud counterparts \cite{dunlop03}). To 
investigate whether ULIRGs go through a QSO phase during their evolution, we 
perform a similar (preliminary) analysis of the kinematics of our sample's 
IR-bright QSOs. We find that the average dispersion of the IR-bright 
QSOs is similar to that of the ULIRG remnants,
indicating that evolutionary links between the two populations may exist. We
intend to expand our study and obtain further spectra of optically 
selected (IR-weak) sources in the future.

\end{abstract}


\section{Introduction}

Galaxy mergers, the frequency of which increases with redshift \cite{kauf93}, 
are considered a key mechanism in driving galactic evolution. 
In the local Universe, the best laboratories for studying violent merging 
events (believed to be the probable analogs to high-redshift mergers) are 
the ultraluminous infrared galaxies (ULIRGs). A plethora of numerical models 
\cite{mihos96},\cite{springel05} and observations \cite{sami96}, 
\cite{kim02} indicates that the ULIRG phase can occur after the first 
encounter of the galaxies and can be present after their nuclei coalesce, 
before complete relaxation sets in.

Several studies indicate that ULIRGs transform gas-rich disks into moderate 
mass ellipticals through merger induced dissipative collapse \cite{mihos96}, 
\cite{kor92}. The structural parameters of a sample of ULIRGs that have 60 
\micron\ flux greater than 1 Jy have been analyzed by Kim et al. (2002) 
and Veilleux et al. (2002), who found that most 
of the sources (73\%) are well-fit by an elliptical-like r$^{1/4}$ light 
profile. Similar findings on the near-infrared (NIR) light distribution 
of ULIRGs was reported by Scoville et al. (2000). 
While the end products of galactic mergers are largely understood, 
the physical details of the merging process (such as the evolution of the host
properties and the black hole growth) are still very uncertain, even in the 
local Universe. It is not 
known, for example, whether ULIRGs (which have a luminosity output $> 10^{12}
L_{\odot}$, comparable to that of QSOS) may go through a QSO phase after the 
nuclear coalescence. 


\section{Observations}

One way to investigate the physical details and the evolution of ULIRGs
is to determine the kinematic and structural properties of the
merging (or interacting) galaxies in different merger phases.  
We are hence conducting an ESO VLT Large Programme (LP; ID 171.B-0442) 
that traces, through NIR spectroscopy, the host dynamics of a sample of 
ULIRGs spanning a wide range of merger phase and infrared (IR) luminosity.
We have obtained high-resolution, H- and K- band ISAAC spectroscopic data. 
Including sources from our previous work \cite{genzel01},\cite{tacconi02} 
we have now observed a total of 53 ULIRGS:
29 of those are merger remnants, 23 are (binary) 
progenitors, and 1 (IRAS 00199-7426) may be a multiple-interaction system 
\cite{duc97}. To investigate whether ULIRGs go through a QSO phase, we have 
also 
acquired ISAAC spectroscopic data for 12 local QSOs, more than half of which 
are IR bright sources (ratio of integrated IR to big blue bump luminosity
$> 0.46$).  

Our on-source integration time varies from 1 to 3 hrs per slit (depending on 
the redshift of the source, [0.046$<$z$<$0.268]), and we use two slits per 
nucleus, so that we can infer its rotation field. We extract  
stellar dispersion $\sigma$ and rotational $V_{\rm rot}$ velocities from 
the spectra using the Fourier correlation quotient 
technique \cite{bender90}; this method provides the line-of-sight (LOS) 
broadening function with which a stellar template has to be convolved to 
produce the observed spectrum.


\section{Pre-coalescence phase}
For the binary (pre-merged) ULIRGs of our sample we are able to derive the
kinematics of the individual progenitors. That allows us to investigate
the conditions that are needed to trigger an ultraluminous burst during a 
merger of gas-rich galaxies. 

\begin{figure}
\vspace*{1.25cm} 
\begin{center}
\epsfig{figure=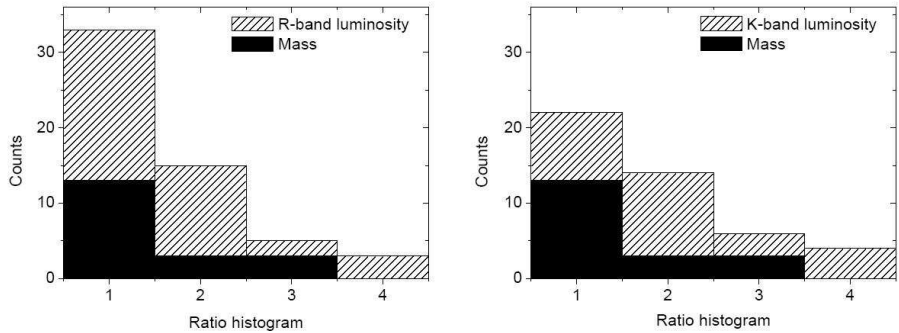,width=13cm} 
\end{center}
\vspace*{0.25cm} 
\caption{ Mass and luminosity ratio histogram. In filled bars we show the mass 
ratio of the ULIRGs in our sample, measured from the stellar kinematics. In 
shaded we show the R-band (left panel) and the K-band (right panel) luminosity 
ratio of the combined 1 Jy and Duc et al. (1997) samples. \label{fig:bins}
} 
\end{figure}

Using the stellar dispersion and, when possible, the stellar rotational 
velocity, we calculate the mass ratio of the merging galaxies. The average 
mass ratio of the binary ULIRGs is 1.5:1, while 68\% of these sources are
1:1 encounters \cite{dasyra05}. This implies that ultraluminous luminosities 
are mainly generated by almost equal-mass mergers. This result is in 
agreement with the luminosity ratios calculated for the 1 Jy sample sources 
\cite{kim02}. In Fig.~\ref{fig:bins} we present a histogram of the merging
galaxies' mass (filled bars) and luminosity (shaded bars) ratios. 
The luminosity ratio distribution is different when measured in the R 
band (left panel) than in the K band (right panel) due to 
stellar population and dust extinction effects. While the overall mass and
luminosity ratio distributions (reasonably) agree, luminosity and mass ratios
of individual sources may deviate. In these cases the mass ratio is the most
robust result because it is less affected by dust extinction and population effects.

We note, however, that we may be undersampling the unequal-mass merger 
categories due to dynamical heating (mainly of the smaller companion) and 
projection effects. We investigate these effects by running 
merger simulations presented in Naab et al. (2003), after adding gas 
(equal to 10\% of the stellar mass). We then 
follow the stellar dispersion as a function of time. We find no significant 
deviations between the apparent and the actual mass ratio of 1:1 mergers, 
while 3:1 mergers may be observed as 2:1 on small nuclear separations. Thus, 
it is likely that only a small fraction of our sample have higher mass ratio 
than measured. 

We can also calculate the sense of rotation of each progenitor using the
(projected) rotational velocity and assuming that the tidal tails are trailing
(whenever archival imaging data allow us to do so). We find that both the
counter-rotating and the less violent, co-rotating, merger geometries may 
lead to ultraluminous activity. 


\section{Post-coalescence phase}
The stellar motions of the coalesced ULIRGs are characterized by an average 
velocity dispersion of 157 \kms\ and a projected rotational velocity of
50 \kms. The velocity dispersion of the late-phase is greater than that of the 
early-phase ULIRGs (142 \kms), which reflects a part of the dynamical heating 
that the merging systems undergo. 

\begin{figure}
\vspace*{1.25cm} 
\begin{center}
\epsfig{figure=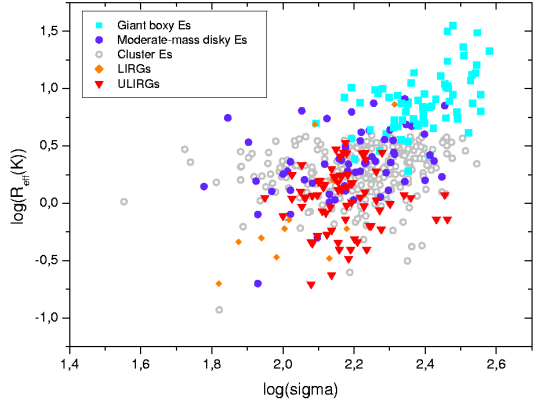,width=9cm} 
\end{center}
\vspace*{0.25cm} 
\caption{The early-type galaxies fundamental plane (R-$\sigma$ projection).
The giant boxy and intermediate-mass disky Es data are taken from 
Bender et al. (1992) and Faber et al. (1997). More (cluster) Es are from 
Pahre (1999) and LIRGs from Shier \& Firscher (1998) and James et al. (1999). 
\label{fig:fpp} 
} 
\end{figure}

When correcting the rotational velocity of the coalesced ULIRGs for
inclination effects (by using an average inclination equal to that of spirals
in the field), we find that the \vrot/$\sigma$ ratio is  
0.64. The ULIRG remnants are, thus, dispersion supported systems, where
rotation is still non-negligible, resembling ellipticals (Es). To investigate
what type of Es can be formed by ultraluminous mergers, Genzel et al. (2001) 
and Tacconi et al. (2002) acquired similar high-resolution NIR spectroscopic 
data for a smaller sample of mostly late merger-stage ULIRGs. By placing their 
sources on the fundamental plane (FP) of early-type galaxies \cite{djoda},  
Tacconi et al. (2002) found indications that ULIRGs resemble intermediate mass 
ellipticals/lenticulars with disky isophotal profiles. From our data we 
can plot the R-$\sigma$ projection of the plane (see Fig.~\ref{fig:fpp}) 
for the sources included in our sample. Giant (boxy) Es are shown in boxes, 
moderate-mass (disky) Es in filled circles, and further cluster Es in open 
cirles. Some luminous infrared galaxies (LIRGs; $10^{11} L_{\odot} < L <
10^{12} L_{\odot}$) are presented in diamonds and the ULIRGs of this study 
in triangles. The fact that ULIRGs clearly populate the intermediate mass  
($\sim$10$^{11}$ M$_{\odot}$) Es part of the FP suggests that these two
populations are linked, while giant Es probably have a different formation 
history.

\section{Implications on the BH growth}

Knowing the stellar dispersion, we can calculate the BH mass, $M_{BH}$, 
with the aid of the $M_{BH}-\sigma$ relation (as presented in Tremaine et al. 
2002; \mbox{$M_{BH}=1.35 \times 10^{8} [\sigma/ 200]^{4.02}$\msun}, 
where $\sigma$ is in \kms ). While the \msigma\ relation 
is valid for virialized systems, it can still be applied to the post-merger
ULIRGs. By the time the two nuclei have coalesced, the host kinematics
have practically reached their relaxation values \cite{beba}. The recent
simulations by Di Matteo et al. (2005) also indicate that merger remnants 
follow this relation. Its application leads to a median remnant black hole 
mass of $8 \times 10^7$\msun. 

On the other hand, the \msigma\ relation needs to be proven valid during the
merger before it can be applied to the dynamically perturbed, non-virialized 
early-phase ULIRGs. Simulations of gas-rich mergers that we ran (see Sect. 3)
show that by the time the merger has advanced to the ULIRG phase, the scatter 
around the \msigma\ relation is small (suggesting that the relation can be 
used). However, these simulations (like those of Di Matteo et al. 2005) assume 
that the amount of gas that accretes onto the black hole from the central 
resolution element of the simulation (denoted as efficiency $\epsilon$) is 
constant. The latter is known to vary drastically when the AGN is switched
on/off. Consequently, any early-phase ULIRG black hole properties derived 
with the aid of the \msigma\ relation are carrying this uncertainty.

\begin{figure}
\vspace*{1.25cm} 
\begin{center}
\epsfig{figure=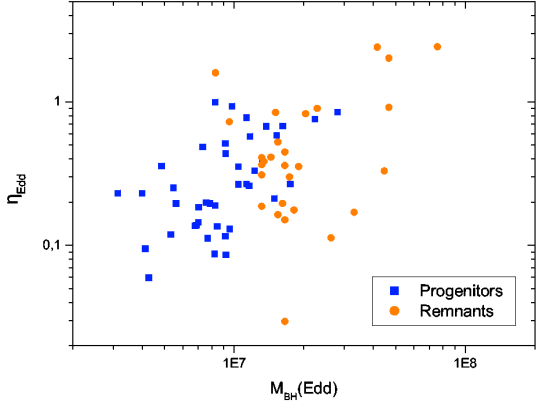,width=9cm} 
\end{center}
\vspace*{0.25cm} 
\caption{ The Eddington efficiency vs the Eddington BH mass of the pre-
and post- coalescence sources. \label{fig:eddi}
}
\end{figure}

We can calculate the Eddington efficiency $\eta_{\rm Edd}$, the ratio between 
the Eddington and the dynamical \mbh, by assigning 50\% of the IR luminosity 
to the AGN. Statistically, this assumption is reasonable given that some 
ULIRGs are largely starburst- while others are AGN- powered \cite{genzel98}.
For individual cases, though, it may make some sources seem to accrete at 
super-Eddington rates; in these cases we assign $\eta_{\rm Edd}$=1.
The Eddington efficiency of the ULIRG remnants is on average 0.52 and it is 
higher than that of the progenitors (see Fig.~\ref{fig:eddi}). 

Photometric study of the 1 Jy sample by Veilleux et al. (2002) indicated 
that the nuclear contribution to the IR luminosity is on average greater 
for the single than for the binary ULIRGs. Thus, the average remnant  
accretion rate we calculated is only a lower limit of its actual value.
Such high $\eta_{\rm Edd}$ values may be an observational confirmation of 
the predictions of Springel et al. (2005) and Di Matteo et al. (2005). These 
authors suggest that after the nuclear coalescence, the gas infall to the 
center of the system is so high that the AGN may even accrete at rates close 
to Eddington.

\section{Discussion: Are there evolutionary links between ULIRGs and QSOs?}

A late-merger evolutionary scenario (originally based on Sanders et al. 1988)
suggests that after the coalescence, the IR emission arising from
the starburst and the nuclear dust is so strong that the remnant reaches
QSO-like luminosities. However, as the dust clears out from the nuclear
region due to AGN winds and supernova feedback, the system  goes through 
an optically bright phase, before further accretion is prevented 
\cite{springel05}.

For the IR-bright sources of our sample reduced thus far, we find that the 
average dispersion is similar to that of the ULIRG remnants. Thus, their black 
hole and host masses are also of comparable size (10$^7$-10$^8$ and 10$^{10}$-
10$^{11}$ \msun\ respectively). The Eddington efficiencies of these QSOs are 
high ($\sim0.5$). Furthermore, the recent HST NICMOS imaging analysis 
of local ULIRGs and IR bright QSOs \cite{veilleux06} indicates that their 
hosts have similar NIR colours. 
 
The IR bright QSOs seem to differ from local QSOs that host supermassive 
black holes, such as the radio-loud (RL) sources or their radio-quiet optical 
counterparts (RQC) of Dunlop et al. (2003). The RL/RQC QSOs have black 
holes of the order 10$^8$ and 10$^9$ \msun\ that accrete on average 
at $\sim$ 0.05 of the Eddington rate, and that are located in $~5$ times
more massive and luminous (K-band) hosts. Probably the radio selected 
QSOs have a different formation history than the IR bright ones.

Our preliminary results seem to indicate evolutionary links between 
ULIRGs and IR bright QSOs; however, the fact that some IR QSOs have
prominent spiral hosts implies that they may have a minor-merger origin. 
We need to improve our statistics and to enrich our sample with optically 
selected QSOs in order to derive conclusions about the optically bright phase
of this scenario.

\section{Conclusions} 

We have acquired spectroscopic H-band, long-slit data of 53 ULIRGs 
(at a variety of merger timescales) and 12 PG QSOs to trace the evolution of 
their dynamical properties. We find that:
\begin{enumerate}

\item
The mass ratio of the binary sources indicates that ULIRGs mainly originate 
from 1:1 and 2:1 mergers. However, less violent events like unequal-mass 
mergers (and possibly co-rotating orientations) are also capable of triggering 
an ultraluminous phase.

\item
Loss of systematic rotation in favor of increase of random motions is
observed from the analysis of the stellar kinematics. The average stellar 
dispersion for the fully merged sources equals 157 \kms. 

\item
The ultraluminous merger will mainly lead to the formation of moderate mass 
ellipticals, similar to those with disky isophotes.  

\item
Using the \msigma\ relation we find that the average \mbh\ of the already
coalesced ULIRGs is of the order 10$^7$-10$^8$ \msun. 

\item
The Eddington efficiency increases from the pre- to the post- merger sources.
The value of the increase is uncertain due to the application of 
the \msigma\ relation for the early-phase (non-virialized) ULIRGs and due 
to the increase of the nuclear luminosity with time. The high efficiency 
of the remnants can be realistic (and attributed to the increase of material 
falling onto the BH). 

\item
The IR bright QSO dispersions and black hole masses, being of the order 
10$^7$-10$^8$ \msun, resemble those of ULIRG remnants, indicating a possible 
link between the two populations. However, they differ from radio-selected
sources that host supermassive black holes of low accretion rates.

\end{enumerate}

\acknowledgements{ We are grateful to N. F\"orster-Schreiber and A. Verma for 
constructive comments. }

\vfill 

\begin{thebibliography}{}{ 

\bibitem{bender90}
Bender, R. 1990,
A\&A, 229, 441

\bibitem{bender92}
Bender, R., Burstein, D., \& Faber, S. M. 1992,
ApJ, 399, 462

\bibitem{beba}
Bendo, G. J., \& Barnes, J. E. 2000,
MNRAS, 316, 315

\bibitem{dasyra05}
Dasyra, K. M., Tacconi, L. J.,  Davies, R.I., Lutz, D., Genzel, R.,
Burkert, A., Veilleux, S. \& Sanders, D. 2005,
submitted to ApJ

\bibitem{dasyra06}
Dasyra, K. M., et al. 2006, in preparation

\bibitem{dimatteo}
Di Matteo, T., Springel, V., \& Hernquist, L. 2005,
2005, Nature, 433, 604

\bibitem{djoda}
Djorgovski, S., \& Davis, M. 1987,
ApJ, 313, 59

\bibitem{duc97}
Duc P.-A., Mirabel, I.F., \& Maza, J. 1997,
A\&AS, 124, 533

\bibitem{dunlop03}
Dunlop, J. S, McLure, R. J., Kukula, M. J., Baum, S. A., O'Dea, C. P., 
\& Hughes, D. H. 2003, MNRAS, 340, 1095

\bibitem{faber97}
Faber, S. M., et al. 1997,
AJ, 114, 1771

\bibitem{genzel98}
Genzel, R., et al. 1998,
ApJ, 498, 579

\bibitem{genzel01}
Genzel, R., Tacconi, L. J., Rigopoulou, D., Lutz, D., \& Tecza, M. 2001,
ApJ, 563, 527

\bibitem{james}
James, P., Bate, C., Wells, M., Wright, G., \& Doyon, R. 1999,
MNRAS, 309, 585 

\bibitem{kauf93}
Kauffmann, G.; \& White, S. D. M. 1993, 
MNRAS, 261, 921

\bibitem{kim02}
Kim, D.-C., Veilleux, S., \& Sanders, D. B. 2002,
ApJS, 143, 277

\bibitem{kor92}
Kormendy, J., \& Sanders, D. B. 1992,
ApJ, 390L, 53

\bibitem{mihos96}
Mihos, J. C., \& Hernquist, L. 1996,
ApJ, 464, 641

\bibitem{naab03}
Naab, T., \& Burkert, A. 2003,
ApJ, 597, 893

\bibitem{pahre}
Pahre, M. A. 1999, 
ApJS, 124, 127

\bibitem{sanders88}
Sanders, D., Soifer, B. T., Elias, J. H., Neugebauer, G., Matthews K. 1988, 
ApJ, 328, L35

\bibitem{sami96}
Sanders, D. B., \& Mirabel, I. F. 1996,
ARA\&A,34,749

\bibitem{sco00}
Scoville, N. Z., Evans, A. S., Thompson, R., Rieke, M., Hines, D. C., Low,
F. J., Dinshaw, N., Surace, J. A., \& Armus, L. 2000
AJ, 119, 991

\bibitem{shier}
Shier, L. M., \& Fischer, J. 1998,
ApJ, 497,163

\bibitem{springel05}
Springel, V., di Matteo, T., \& Hernquist, L. 2005,
MNRAS, 361, 776

\bibitem{tacconi02}
Tacconi, L. J., Genzel, R., Lutz, D., Rigopoulou, D., Baker, A. J., Iserlohe,
C., \& Tecza, M.  2002,
ApJ, 580, 73

\bibitem{tremaine02}
Tremaine, S., et al. 2002,
ApJ, 574, 740

\bibitem{veilleux02}
Veilleux, S., Kim D.-C., \& Sanders, D. B. 2002,
ApJS, 143, 315

\bibitem{veilleux06}
Veilleux, S., et al. 2006, in preparation

} \end{thebibliography}
\end{document}